# High responsivity phototransistors based on few-layer ReS$_2$ for weak signal detection

*Erfu Liu[1], Mingsheng Long[1], Junwen Zeng[1], Wei Luo[1,2], Yaojia Wang[1], Yiming Pan[1], Wei Zhou[1], Baigeng Wang[1\*], Weida Hu[3], Zhenhua Ni[4], Yumeng You[5], Xueao Zhang[2], Shiqiao Qin[2], Yi Shi[6], K. Watanabe[7], T. Taniguchi[7], Hongtao Yuan[8,9\*], Harold Y. Hwang[8,9], Yi Cui[8,9], Feng Miao[1\*] and Dingyu Xing[1]*

**Two-dimensional transition metal dichalcogenides are emerging with tremendous potential in many optoelectronic applications due to their strong light-matter interactions. To fully explore their potential in photoconductive detectors, high responsivity is required. Here, we present high responsivity phototransistors based on few-layer rhenium disulfide (ReS$_2$). Depending on the back gate voltage, source drain bias and incident optical light intensity, the maximum attainable photoresponsivity can reach as high as 88,600 A W$^{-1}$, which is a record value compared to other individual two-dimensional materials with similar device structures and two orders of magnitude higher than that of monolayer MoS$_2$. Such high photoresponsivity is attributed to the increased light absorption as well as the gain enhancement due to the existence of trap states in the few-layer ReS$_2$ flakes. It further enables the detection of weak signals, as successfully demonstrated with weak light sources including a lighter and limited fluorescent lighting. Our studies underscore ReS$_2$ as a promising material for future sensitive optoelectronic applications.**

[1] National Laboratory of Solid State Microstructures, School of Physics, Collaborative Innovation Center of Advanced Microstructures, Nanjing University, Nanjing 210093, China.

[2] College of Science, National University of Defense Technology, Changsha 410073, China.

[3] National Laboratory for Infrared Physics, Shanghai Institute of Technical Physics, Chinese Academy of Sciences, Shanghai 200083, China.

[4] Department of Physics, Southeast University, Nanjing 211189, China.

[5] Department of Chemistry, Southeast University, Nanjing 211189, China.

[6] School of Electronic Science and Engineering, Nanjing University, Nanjing 210093, China.

[7] National Institute for Materials Science, 1-1 Namiki, Tsukuba, 305-0044, Japan.

[8] Geballe Laboratory for Advanced Materials, Stanford University, Stanford, California 94305, USA.

[9] Stanford Institute for Materials and Energy Sciences, SLAC National Accelerator Laboratory, Menlo Park, California 94025, USA.

Correspondence and requests for materials should be addressed to F. M. (E-mail: miao@nju.edu.cn), H. T. Y. (E-mail: htyuan@stanford.edu), or B. W. (E-mail: bgwang@nju.edu.cn).

## 1. Introduction

As the most important class of semiconducting 2D (two-dimensional) materials, the family of transition metal dichalcogenides (TMDs)[1,2] (MX$_2$ where M denotes a transition metal and X denotes a chalcogen) have shown promising properties for optoelectronic[3-8] applications due to strong light-matter interaction. For example, MoS$_2$ has been demonstrated to be a suitable material for building a sensitive phototransistor[4,9], which is the most fundamental and key component of many optoelectronic circuits. To fully realize the potential application of TMD materials in photodetection, high responsivity is required. The photoresponsivity $R_\text{p}$ of a photoconductive detector is equal to the product of the intrinsic responsivity $R_0$ and the photoconductive gain $G$ and can be expressed by

$$R_\text{p} = R_0 \cdot G = \frac{\eta \text{e}}{h\upsilon} \cdot G \qquad (1)$$

where $\eta$ is the quantum efficiency, $e$ is electron charge, $h$ is the Planck constant, and $\upsilon$ is the frequency of the light. Semiconducting materials with direct bandgaps are of primary interest due to efficient light absorption, resulting in high quantum efficiency. Many studies of phototransistors have focused on monolayer Mo- and W-based transition metal dichalcogenides[3,4,8,10-12] with direct bandgaps[1,13]. However, due to various limitations, the reported photodetection efficiency has been low[3,5,11,14-16]. Thus, searching for new materials with high responsivity is highly desirable.

As a new member of TMDs, rhenium disulfide (ReS$_2$) has been discovered to be a direct band gap semiconductor due to weak interlayer coupling[17]. ReS$_2$-based high performance field effect transistors[18,19] and digital devices[19] have also been recently demonstrated. Taking advantage of its direct band gap property, few-layer ReS$_2$ with larger thickness could have a higher density of states[20] and enhanced light absorption[1], potentially making it an ideal candidate to realize high responsivity and weak signal detection.

In this work, we demonstrated high photoresponsivity based on few-layer ReS$_2$ phototransistors. The optimized photoresponsivity reached as high as 88,600 A W$^{-1}$ at a wavelength of 532 nm, which is 100 times better than that of monolayer MoS$_2$[4] and more than 5,000-fold improvement over the previously reported results[21]. Although 2D materials decorated with quantum dots[22] or stacked van der Waals heterostructures[23] could reach higher photoresponsivity, our result is still a record value compared to all other individual 2D materials-based phototransistors with similar two-terminal structures[2,4,5,15,24-31]. The high photoresponsivity of few-layer ReS$_2$ photodetectors originates from increased light absorption due to larger film thickness and efficient gain enhancement due to the existence of trap states. Taking advantage of the high photoresponsivity of the few-layer ReS$_2$ photodetectors, we also successfully demonstrated the detection of very weak signals by using a lighter and limited fluorescent lighting in a dark room as the weak light sources. Our results suggest few-layer ReS$_2$ as a promising material for future sensitive optoelectronic applications.

## 2. Results and Discussion

**2.1 Fabrication and characterizations of few-layer ReS$_2$ phototransistors.**

**Figure 1**a shows the schematics of a few-layer ReS$_2$ phototransistor. Figure 1b and 1c show the typical optical microscopy and atomic force microscopy (AFM) images of a phototransistor device (the Raman characterization results are presented in Supporting Information). The substrate under the h-BN layer is a standard silicon wafer covered by a 285-nm-thick oxide layer. The thickness of the few-layer ReS$_2$ was determined by AFM. In this study, we focused on few-layer ReS$_2$ flakes, with thicknesses ranging from 2.5 nm to 4.5 nm (3-6 layers with interlayer spacing of approximately 0.7 nm). To reduce the scattering from impurities in the substrate, we used h-BN rather than SiO$_2$ as the substrate; h-BN has been shown to improve the mobility in graphene[32] and MoS$_2$[33].

We first performed the electrical characterizations of the ReS$_2$ phototransistors in the dark state using the measurement circuit shown in Figure 1a. All measurements in this work were performed at room temperature and under ambient conditions. **Figure 2**a shows the transfer curve of a typical device (Dev#10) with a semi-log plot of the source-drain current ($I_{ds}$) versus back gate voltage ($V_{bg}$) in the dark state. The source-drain voltage ($V_{ds}$) was fixed at 1.0 V. The device showed excellent *n*-type field effect transistor (FET) behaviors, where the on/off ratio reached $10^8$ and the off-state current was smaller than 1 pA. The current approached saturation when $V_{bg}$ reached 60 V. The threshold voltage $V_t$ was approximately -30 V, indicating natural *n*-doping or large electron concentration in the device, which was observed in most of our devices. The *n*-doping may be attributed to the impurities or S vacancies in the $ReS_2$ flakes. The field-effect mobility can be determined by taking the steepest slope in the transfer curves (see the Experimental Section for the calculation details). The mobility of all few-layer $ReS_2$ devices we measured varied between 5 and 30 $cm^2$ $V^{-1}$ $s^{-1}$, and the highest mobility we obtained was in a 6-layer device.

**2.2 Photoresponse of few-layer ReS₂ photodetectors.**

We further measured the transfer characteristics under light illumination. Because the direct band gap of $ReS_2$ ranges from 1.58 eV (monolayer) to 1.5 eV[17] (bulk), corresponding to photon wavelengths from 785 nm to 826.7 nm, we used a laser with a 532 nm wavelength to generate a photocurrent. The light was provided by a focused laser beam with a spot diameter less than 1 μm (see more details in Supporting Information). During the measurements, the power of the incident light $P_{light}$ and $V_{ds}$ were kept at 20 nW and 1.0 V, respectively. Figure 2b shows the linear scale transfer characteristics of both the dark and illuminated states (dev#12). The current under illumination was obviously enhanced over the entire range of $V_{bg}$ from -60 V to +60 V. We then calculated the photoresponsivity $R_p$, which is a crucial parameter for the performance of a photodetector. It is defined as

$$R_p = I_p/P_{light} \quad (2)$$

where $I_p$ is the generated photocurrent ($I_p = I_{illuminated} - I_{dark}$) and $P_{light}$ is the power of the incident illumination. The inset in Figure 2b shows the calculated photoresponsivity as a function of $V_{bg}$. When $V_{bg}$ was approximately 0 V, the photoresponsivity reached the maximum value, approximately 1067 A $W^{-1}$. It monotonically decreased when $V_{bg}$ increased to 60 V or decreased to -60 V.

This back gate voltage-dependent photoresponsivity can be readily explained by effective tuning of the carrier density of the channel and Schottky barriers near the electrodes in the ReS$_2$ phototransistors. Without applying a source drain bias or back gate voltage in the dark state, due to the natural *n*-doping of the ReS$_2$ flakes the Fermi level $E_F$ is close to the conduction band edge ($E_C$), as schematically shown in Figure 2c. Schottky barriers formed at the interfaces between the Ti/Au electrodes and the ReS$_2$. Applying a source-drain bias effectively drives the photogenerated carriers between the electrodes, as shown in Figure 2d. Thus, illuminating the device produces highly efficient photocurrent extraction, resulting in a large photoresponsivity. When turning off the device by decreasing $V_{bg}$ (when $V_{bg} < V_t$), as shown in Figure 2e, $E_F$ falls into the band gap, resulting in an increase in the Schottky barrier height between the electrodes and the ReS$_2$ and further suppression of the generated photocurrent. By increasing $V_{bg}$, the device approaches the saturated state and $E_F$ moves into the conduction band (as shown in Figure 2f). The total number of the excited states is fixed with the 532 nm laser pumping. Approaching the saturated state indicates that most of the states in the conduction band are filled. Thus, a majority of photogenerated electron-hole pairs recombine very fast. The number of photogenerated carriers which contribute to the current is greatly reduced, resulting in a decreased photoresponsivity.

**Figure 3**a shows the photoswitching characteristics of a few-layer ReS$_2$ phototransistor with the bias voltage in the range of 1.0 V to 4.0 V. $V_{bg}$ was fixed at -50 V to reduce the dark current, and the incident laser power was fixed at 20 nW. The switching behavior of the channel current was clearly observed, with the response time ranging from a few to tens of seconds, with similar results reported in hybrid graphene-quantum dot[22] or monolayer MoS$_2$ photodetectors[4]. The current remained low when the device was in the dark state and increased to a much higher value after illumination from the laser. After stopping the illumination, the current dropped back to the low value. The calculated photoresponsivity as a function of the bias voltage is shown in Figure 3b, with the value increasing from 61 A W$^{-1}$ to 2515 A W$^{-1}$ when the bias voltage increased from 1.0 V to 5.0 V.

We further investigated the photoresponsivity under different excitation powers.

The inset in Figure 3c displays the measured current as a function of $V_{ds}$ with various $P_{light}$. The photocurrent increased gradually with increasing incident optical power. Figure 3c shows the photocurrent under illumination as a function of the incident power when the bias was fixed at 4.0 V and the back gate voltage was fixed at -50 V. For $P_{light}$ ranging from 6 pW to 29 μW, the photocurrent had a sub-linear relationship with the optical power, similar to the results reported in other 2D materials[4,15,34,35]. Figure 3d presents the measured photoresponsivity versus the incident optical power, as well as the results from a few representative 2D materials for comparison. When the incident optical power approached 6 pW, the measured photoresponsivity of our few-layer ReS$_2$ phototransistor reached as high as 88,600 A W$^{-1}$ and the external quantum efficiency reached $2\times10^7$%. To the best of our knowledge, this is the highest reported photoresponsivity for phototransistors based on 2D materials with similar two-terminal structures[5,15,25,26] (the comparison of photodetectors based on different 2D materials is presented in Supporting Information). This result is also more than 100 times better than the results reported for monolayer MoS$_2$ phototransistors[4] (880 A W$^{-1}$) and over $10^7$ times higher than the first graphene photodetector[14] (approximately 0.5 mA W$^{-1}$). We also calculated the noise equivalent power (NEP) and Specific detectivity ($D^*$). When the incident optical is 6 pw ($V_{bg} = -60$ V and $V_{ds} = 4$ V), NEP $= 1.277 \times 10^{-18}$ W Hz$^{-1/2}$ and $D^* = 1.182 \times 10^{12}$ Jones ($D^* = RA^{1/2}(2eI_d)^{-1/2}$, NEP=$(2eI_d)^{1/2}R^{-1}$ where $R$=88600 AW$^{-1}$ is the responsivity, $A$=2.28 μm$^2$ is the area of sample, e is value of electron charge, and $I_d$=40 nA is the dark current). This specific detectivity is close to the value of silicon photodetectors[29] (~4×10$^{12}$ Jones).

**2.3 Enhancement mechanism for high photoresponsivity.**

As described in equation (1), the photoresponsivity of a photodetector is equal to the product of the intrinsic responsivity $R_0$ and the photoconductive gain $G$. The high photoresponsivity in the few-layer ReS$_2$ phototransistors, up to 88,600 A W$^{-1}$, is induced by the enhancement of both quantum efficiency $\eta$ and gain $G$.

The quantum efficiency $\eta$ is proportional to the absorbed optical power $P_d$, which is directly related to the film thickness and defined by:

$$P_d = P_0(1 - e^{-\alpha d}) \qquad (3)$$

where $P_0$ is the incident optical power, $\alpha$ is the absorption coefficient and $d$ is the film thickness. For multiple-layer ReS$_2$ films, due to the advantage of larger thickness, $P_d$ and thus $\eta$ are notably increased[1,36].

Regarding the enhancement of $G$, it has been comprehensively discussed in many similar systems, and in our case[37], it is likely attributed to the existence of trap states. The trap states could be mainly contributed by two sources, including the impurities or S vacancies in the ReS$_2$ flakes, and the surface contamination due to ambient species or fabrication processes. As shown in **Figure 4**a, under illumination, the electrons excited from the valance band fill some of the trap states and remain. Due to the charge conservation in the channel, the holes in the valance band are continuously replenished by the drain electrode when the same amount of holes reaches the source electrode. Multiple holes circulate after a single photon generates an electron-hole pair. Due to the long times that the electrons remain in the trap states, the lifetimes of the excited holes $T_{\text{lifetime}}$ in ReS$_2$ are very long, which leads to a greatly enhanced gain according to $G = T_{\text{lifetime}}/T_{\text{transit}}$[38]. We also estimated the density of the trap states in ReS$_2$ devices by studying the temperature dependence of the field-effect mobility (for details see Supporting Information), which gave the value of the trap state density $n_{trap}$ of $1.96 \times 10^{13}$ cm$^{-2}$.

**2.4 Weak signal detection.**

Due to the high photoresponsivity achieved in our ReS$_2$ phototransistors, they may be suitable for the detection of weak signals, which has broad applications. As a demonstration, we used a 5-layer ReS$_2$ phototransistor and a lighter and limited fluorescent lighting in a dark room as the weak light sources. The back gate voltage was fixed at -50 V and the source drain bias was fixed at 2.0 V. The photoresponse curve is shown in Figure 4b, where the photocurrent reached approximately 1 nA and 7.6 nA under the illumination of a lighter and limited fluorescent lighting, respectively.

## 3. Conclusions

In conclusion, we successfully fabricated high responsivity phototransistors based on few-layer ReS$_2$. By optimizing the device test parameters, a record high responsivity, 88,600 A W$^{-1}$ was achieved, which is higher than other individual 2D material based photodetectors with similar two-terminal structures. Further investigations revealed that the high responsivity originates from the increased light absorption and gain enhancement due to the existence of trap states in the few-layer ReS$_2$ flakes. Taking advantage of the high photoresponsivity of the few-layer ReS$_2$ photodetectors, we successfully demonstrated the detection of weak signals using a lighter and limited fluorescent lighting in a dark room as the weak light sources. Our results point to few-layer ReS$_2$ with high photoresponsivity as a promising material for future optoelectronic applications.

During the preparation of this manuscript, we became aware of another work studying the few-layer black phosphorus-based photodetectors[39], for which the photoresponsivity is the same order as our few-layer ReS$_2$ based photodetectors.

## 4. Experimental Section

*Materials and devices:* Single crystals of ReS$_2$ were purchased from 2Dsemiconductors.com. We used a standard mechanical exfoliation method to isolate few-layer ReS$_2$ flakes. Few-layer ReS$_2$ films were transferred onto thin hexagonal boron nitride (h-BN) using polymethylmethacrylate (PMMA) as an intermediate membrane[32]. The number of layers could be initially estimated by observing the color interference on a 285-nm-thick SiO$_2$ wafer and further confirmed by measuring the thickness with a Bruker Multimode 8 AFM. A conventional electron-beam lithography process followed by standard electron-beam evaporation of metal electrodes (typically 5 nm Ti/40 nm Au) was used to fabricate few-layer ReS$_2$ phototransistors.

*Calculation details of the device mobility:* The field-effect mobility can be calculated using the equation

$$\mu = \frac{Ld}{WV_{\text{ds}}\varepsilon_0\varepsilon_r}\frac{dI_{\text{ds}}}{dV_{\text{bg}}}$$

where L and W are the channel length and width, respectively, $\varepsilon_0$ is the vacuum permittivity, $\varepsilon_r$ for SiO$_2$ is 3.9, and the thickness $d$ of the SiO$_2$ is 285 nm. Here, we ignored the capacitance of the thin h-BN (the h-BN we used was approximately 14 nm thick and contributed to a minor decrease in the calculated mobility). The mobility calculated from the transfer curve of the inset of Figure 2a was approximately 27 cm$^2$ V$^{-1}$ s$^{-1}$.

**Supporting Information**

Supporting Information is available from the Wiley Online Library or from the author.

**Acknowledgements**

This work was supported in part by the National Key Basic Research Program of China (2015CB921600), the Natural Science Foundation of Jiangsu Province (BK20140017), the National Key Basic Research Program of China (2013CBA01603), the National Natural Science Foundation of China (11374142, 11322441, 61574076), the Natural Science Foundation of Jiangsu Province (BK20130544, BK20150055), the Specialized Research Fund for the Doctoral Program of Higher Education (20130091120040), and Fundamental Research Funds for the Central Universities and the Collaborative Innovation Center of Advanced Microstructures. H. T. Y., H. Y. H., and Y. C. were supported by the Department of Energy, Office of Basic Energy Sciences, Division of Materials Sciences and Engineering, under contract DE-AC02-76SF00515. F. M. and E. L. thank Dr. Zheng Liang at the Graphene Research and Characterization Centre and Taizhou Sunano New Energy Co., Ltd. for excellent experimental assistance.

F. M. and E. L. conceived the project and designed the experiments. E. L., M. L. and J. Z. carried out the device fabrication and electrical measurements. E. L., F. M. and H. T. Y. performed the data analysis and interpretation. F. M., E. L. and H. T. Y. co-wrote the paper, with all authors contributing to the discussion and preparation of the manuscript.

**Competing financial interests**

The authors declare that they have no competing financial interests.

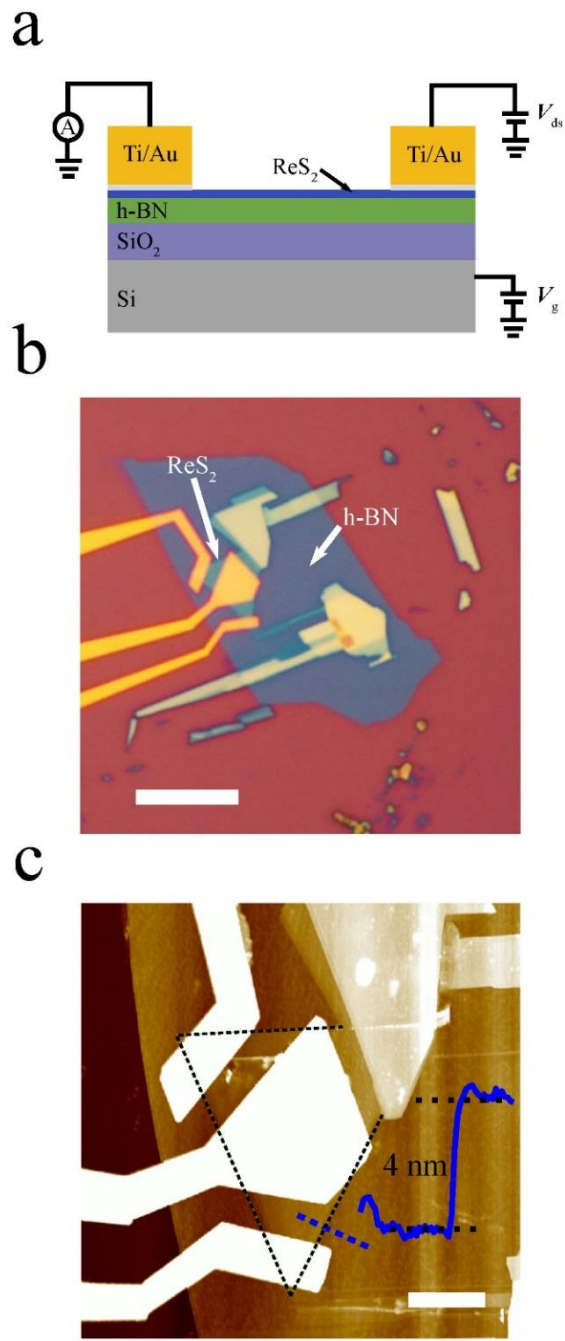

**Figure 1. Few-layer ReS$_2$ phototransistors.** (**a**) Schematic and measurement circuit of the few-layer ReS$_2$ phototransistors. (**b**) Optical microscopy image of a few-layer ReS$_2$ phototransistor. The scale bar is 10 μm. (**c**) AFM image of the same device. The scale bar is 2 μm. The height of the ReS$_2$ flake is approximately 4.0 nm, indicating the number of layers to be 5.

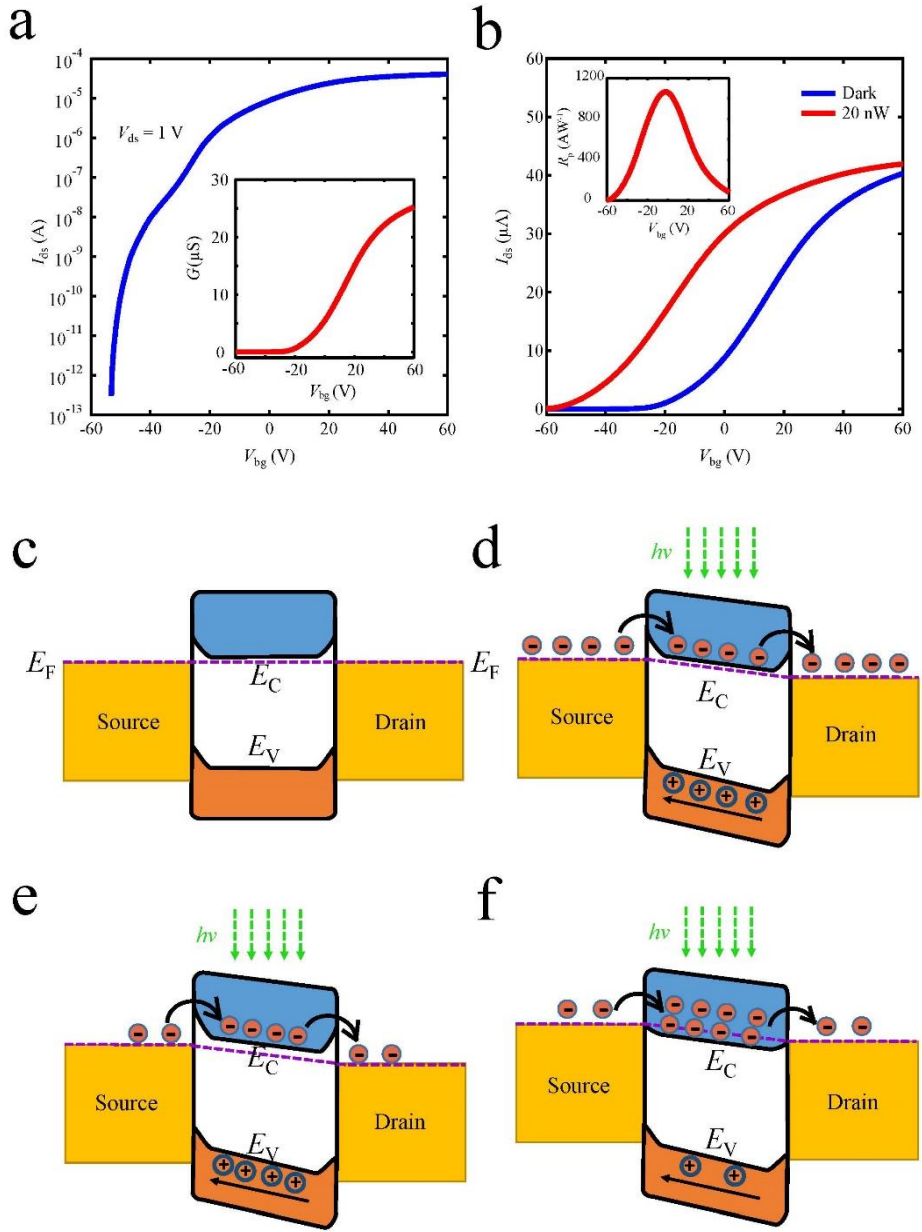

**Figure 2. Characterization of few-layer ReS$_2$ phototransistors.** (**a**) The transfer curve of a ReS$_2$ transistor. $V_{ds}$ was fixed to 1.0 V. The on/off ratio reached $10^8$. Inset: device conductance as a function of $V_{bg}$. (**b**) Linear transfer curves in the dark state and under illumination. $P_{light}$ was 20 nW. Inset: photoresponsivity ($R_P$) as a function of $V_{bg}$. (**c**) Band diagram under open circuit without illumination. (**d**) Band diagram of the biased ON state under illumination. (**e**) Band diagram of the biased OFF state under illumination. (**f**) Band diagram of the biased state approaching the saturated states under illumination. $E_F$, $E_C$ and $E_V$ represent the Fermi level, conduction band edge and valence band edge, respectively.

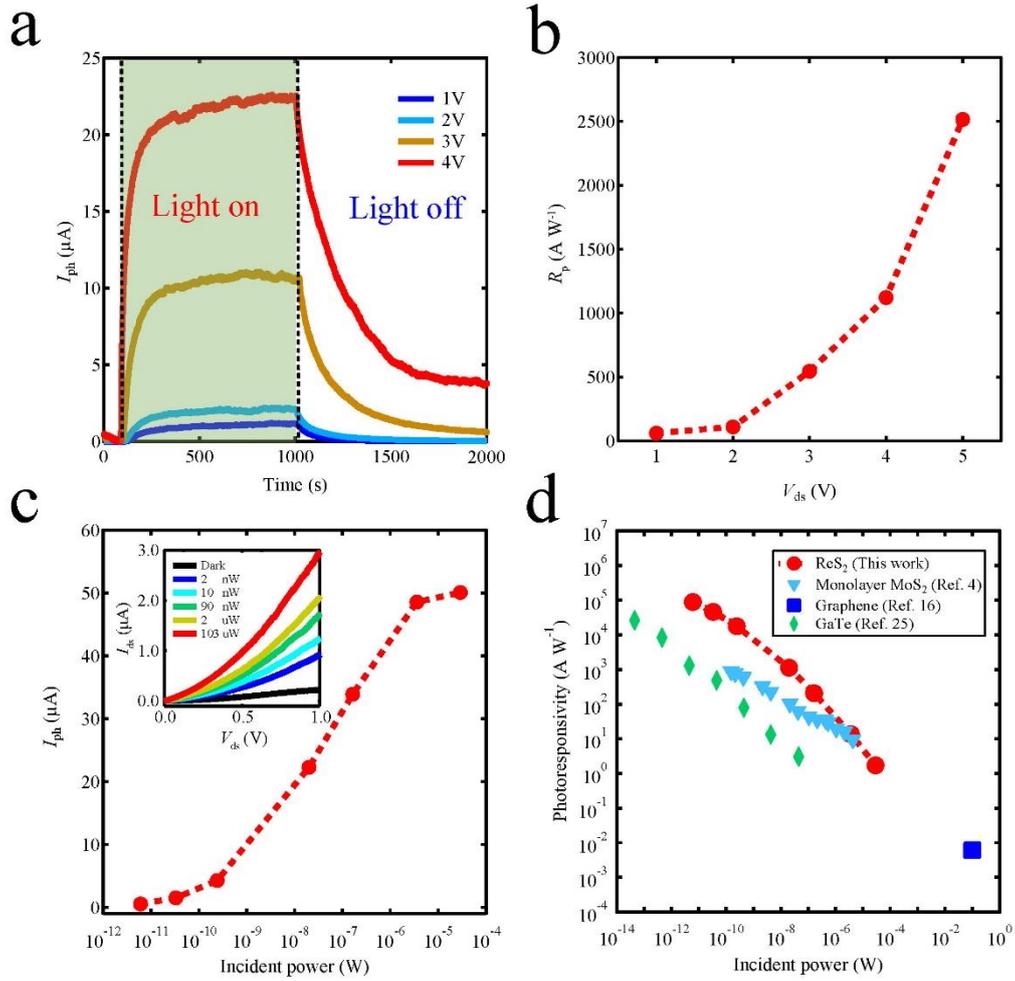

**Figure 3. Photoresponse with high photoresponsivity.** (**a**) Photoswitching behaviors under various $V_{ds}$ with $P_{light}$ = 20 nW and $V_{bg}$ = -50 V. (**b**) The dependence of $R_P$ on $V_{ds}$. (**c**) The dependence of photocurrent on $P_{light}$ with $V_{ds}$ = 4.0 V and $V_{bg}$ = -50 V. Inset: $I_{ds}$-$V_{ds}$ curves of the ReS$_2$ phototransistor under various $P_{light}$. (**d**) The dependence of $R_P$ on $P_{light}$. Under weak illumination (approximately 6 pW), $R_P$ reached as high as 88,600 A W$^{-1}$. For comparison, the results from monolayer MoS$_2$, multilayer GaTe and graphene are also presented. Error bar is smaller than the size of the dots.

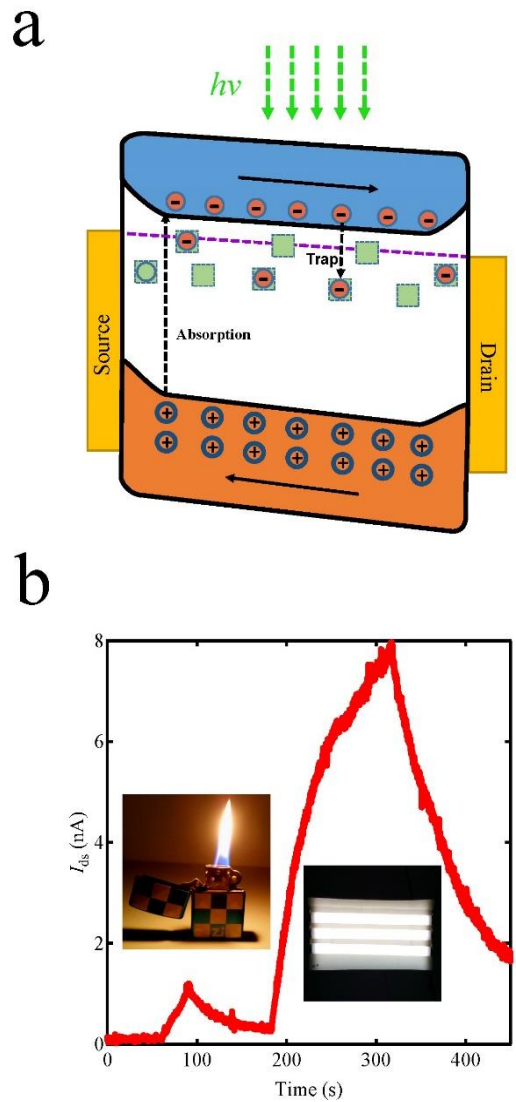

**Figure 4. Gain enhancement and weak signal detection.** (**a**) Schematic showing the enhancement of the photoconductive gain due to the existence of trap states in ReS$_2$. Under illumination, multiple holes circulate after a single photon generates an electron-hole pair. The purple dash line represents the Fermi level. (**b**) Weak signal detection in a 5-layer ReS$_2$ phototransistor using a lighter and limited fluorescent lighting as the weak light sources. $V_{bg}$ was fixed at -50 V and $V_{ds}$ was fixed at 2.0 V. The areas with light green background represent the states under illumination.

# Supporting Information

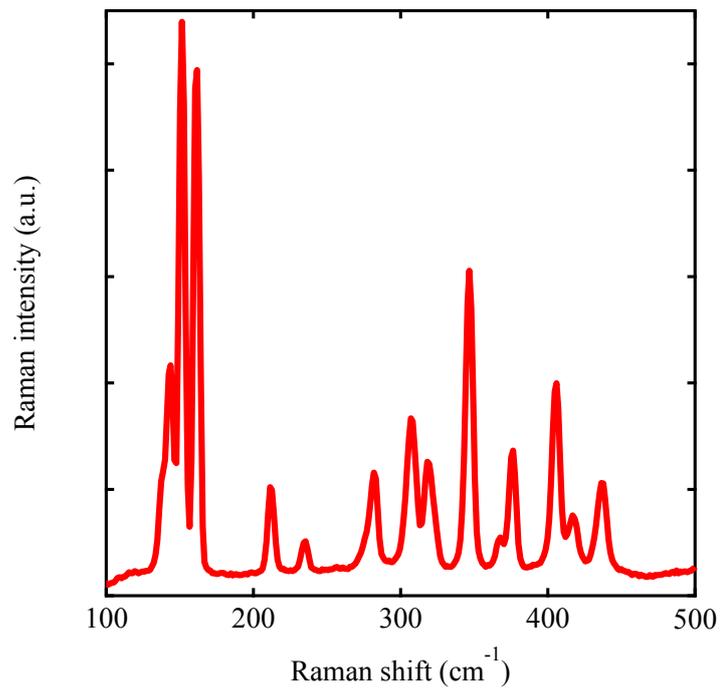

**Figure S1: Raman spectrum of a ReS$_2$ phototransistor.** Due to the low symmetry lattice structure of ReS$_2$, more Raman modes were observed than in other TMD materials. More details can be found in our recent work [1].

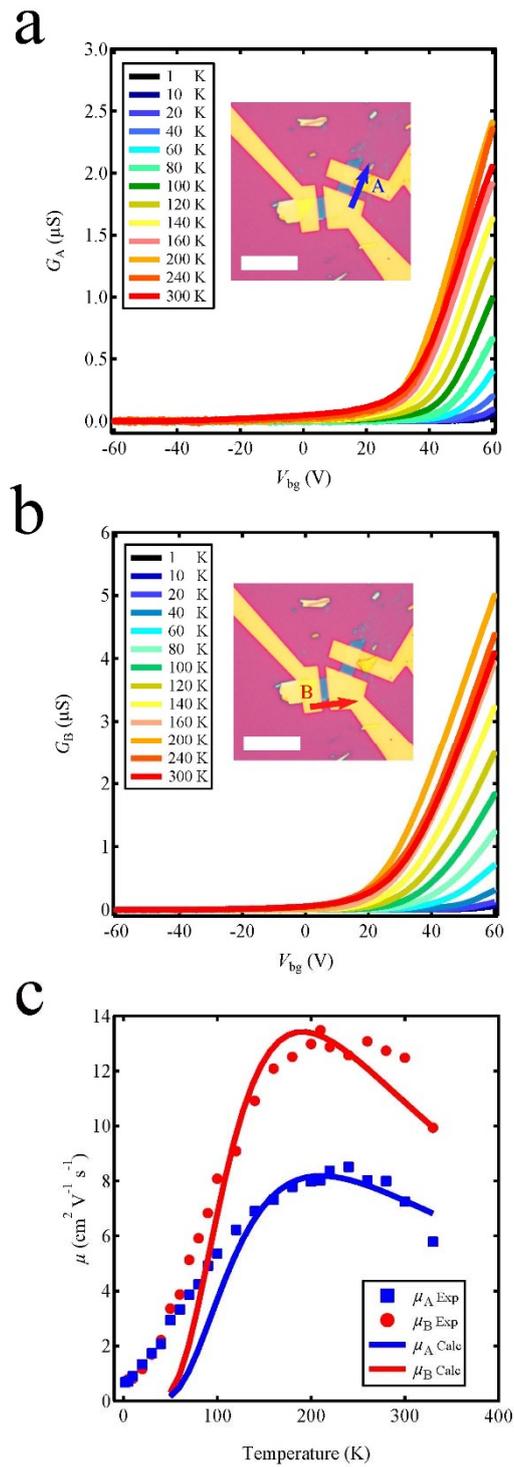

**Figure S2: Temperature dependence of the field-effect mobility.** a) Transfer curves from 1.6 K to 300 K along the A-direction of a few-layer ReS$_2$ FET. The inset is the optical image of the device. The scale bar is 10 μm. b) Transfer curves from 1.6 K to 300 K along the B-direction. b) Mobility (μ) versus temperature along the two directions.

**The measurement of the incident light power.**

During our measurements, when the light power was greater than 1 nW, we used a focused laser beam to provide the illumination. The laser power ranged from 1 nW to approximately 5 mW. Because the spot size of the focused beam (with diameter less than 1 μm) is much smaller than the device channel size, the incident optical power was equal to the total laser power.

When the light power was less than 1 nW, due to the limitations of commercial power meters, we used a non-focused laser beam instead. The mean optical power per unit area $P_{\text{mean}}$ could be calculated by assuming a spot size radius of approximately 3 mm, and the total optical power of the laser $P_{\text{total}}$ was measured using a Newport 1918-R optical power meter. The optical power illuminated on the devices was then determined by $P_{\text{in}} = P_{\text{mean}} \times A$, where A is the area of the devices.

**Temperature dependence of the field-effect mobility.**

Because the low lattice symmetry induced anisotropic properties in ReS$_2$, we measured two field effect transistors (FETs) along two principle axes (see the inset of Supplementary Figure 2c). Using the method described in our previous work [2], the two directions of the flake were determined to be the a- and b-axes (denoted A and B, respectively). The FET transfer curves along the two directions at different temperatures (from 1.6 K to 330 K) are shown in Figure S2a and S2b, respectively. After extracting the mobility values from the transfer curves around the carrier density $n = C_{\text{bg}} V_{\text{bg}} = 4 \times 10^{12} \text{cm}^{-2}$ (where $C_{\text{bg}}$ = 12.2 nF cm$^{-2}$ for 285-nm-thick SiO$_2$ dielectric and $V_{\text{bg}}$ = 55 V), we obtained the dependence of the field-effect mobility on the temperature in the two directions (Figure S2c).

The experimentally measured mobility of ReS$_2$ could have been affected by a few major extrinsic factors, such as charge traps [3], charge impurities [4, 5] and short-range scattering [6]. In fact, our results on the temperature-dependent transport properties of ReS$_2$ can be quantitatively understood by adapting the multiple trapping and release model. As shown in Figure S2c, the experimental data at high temperatures (150 K-300 K) can be fitted well with this model. At low temperatures (<100 K), however, the calculated $\mu$ is lower than experimental result, presumably due to the omission of hopping transport between traps in our model. The hopping transport usually does not have a strong dependence on transport directions, hence the suppression of the anisotropic mobility ratio at low temperatures. Our results indicate that the device mobility is not dominated by phonon scattering but rather limited by some extrinsic factors, especially charge traps.

**The multiple trapping and release model**

To estimate the density of the charge trap states in the few-layer ReS$_2$ flakes, we assumed the charge traps were uniformly distributed within a trap distribution width $\Delta E_{tr}$ below the conduction band edge. Here, we only considered shallow traps because the density of the deep traps does not affect the mobility and conductivity [7, 8]. The Fermi energy $E_F(n, T)$ is determined by

$$n = C_g V_g = \int_0^{+\infty} N_0 \frac{1}{e^{(E-E_F)/k_B T} + 1} dE + \int_{-\Delta E_{tr}}^{0} \frac{N_{tr}}{\Delta E_{tr}} \frac{1}{e^{(E-E_F)/k_B T} + 1} dE$$

where $N_{0A} = \frac{2m_A^*}{\pi \hbar^2} = 9.4 \times 10^{14}\ eV^{-1} cm^{-2}$, $N_{0B} = \frac{2m_B^*}{\pi \hbar^2} = 5.0 \times 10^{14}\ eV^{-1} cm^{-2}$ are the densities of states in the conduction band. Here, according to our *ab initio* calculations, $m_A^* = 1.11\ m_e$ and $m_B^* = 0.59\ m_e$ are the conduction band effective masses along the A- and B-directions, respectively. The density of conducting electrons in the extended states is

$$n_c(n, T) = \int_0^{+\infty} N_0 \frac{1}{e^{(E-E_F)/k_B T} + 1} dE$$

Here, we assumed the carriers occupying the trap states do not carry current. Therefore, the "effective" mobility $\mu$ is given by

$$\mu(n, T) = \mu_0(T) \frac{\partial n_c(n, T)}{\partial n}$$

where $\mu_0$ is the free-carrier mobility, which is simply assumed to have a temperature dependence of $\mu_0(T) = \alpha T^{-2}$ by considering phonon-limited scattering. In Supplementary Figure 3b, the fitting parameters of shallow traps we obtained include the trap state density $N_{tr} = 1.96 \times 10^{13} \ cm^{-2}$, $\Delta E_{tr} = 49 \ meV$, and the corresponding value for the room temperature $\mu_0$ at room temperature ($= \alpha/(300 \ K)^2$) are $= 29.8 \ cm^{-2}/Vs$.

**Comparison of photodetectors based on 2D materials.**

We compared the performance of photodetectors based on semiconducting layered materials with similar 2-terminal or 3-terminal phototransistor structures in Table S1.

**Table S1:** Comparison of photodetectors based on 2D materials.

| Material | Device structure | Responsivity (A W$^{-1}$) | Response time | Spectral range | Reference |
|---|---|---|---|---|---|
| ReS$_2$ | 3 terminal | 88,600 | Tens of seconds | Visible | This work |
| Graphene | 3 terminal | 0.01 | 1.5 ps | Visible - IR | 9 |
| Single-layer MoS$_2$ | 3 terminal | 880 | 2 s | Visible | 10 |
| Single-layer MoS$_2$ | 3 terminal | 7.5×10$^{-3}$ | 50 ms | Visible | 11 |
| Multilayer MoS$_2$ | 3 terminal | 0.1 | >1 s | Visible - IR | 12 |
| Multilayer WS$_2$ | 2 terminal | 92×10$^{-6}$ | 5.3 ms | Visible | 13 |
| Multilayer GaSe | 2 terminal | 2.8 | 20 ms | UV - Visible | 14 |
| Multilayer GaTe | 3 terminal | 10$^4$ | 6 ms | Visible | 15 |
| Multilayer GaS | 2 terminal | 4.2 | 30 ms | UV - Visible | 16 |
| Multilayer In$_2$Se$_3$ | 2 terminal | 395 | 18 ms | UV - IR | 17 |
| Multilayer b-P | 3 terminal | 0.0048 | 1 ms | Visible - IR | 18 |